\documentclass[pre,amsmath]{revtex4}
\usepackage{graphicx}
\usepackage{color}
\usepackage{bm}
\usepackage{epsfig}
\usepackage{latexsym}
\usepackage{pifont}
\usepackage{float}
\usepackage[utf8]{inputenc}
\graphicspath{{figure/}}
\usepackage{siunitx}

\begin{document}

\newcommand{\be}{\begin{equation}}
\newcommand{\ee}{\end{equation}}
\newcommand{\bea}{\begin{eqnarray}}
\newcommand{\eea}{\end{eqnarray}}
\newcommand{\nn}{\nonumber}

\title{Interplay between surface and bending energy helps membrane protrusion formation }
\author{Raj Kumar Sadhu\footnote{rajkumar.sadhu@bose.res.in} and Sakuntala Chatterjee}
\affiliation{Department of Theoretical Sciences, S. N. Bose National Centre for Basic Sciences, Block  JD, Sector  III, Salt Lake, Kolkata  700106, India.}

\begin{abstract}
We consider a one-dimensional elastic membrane, which is pushed by growing filaments. The filaments tend to grow by creating local protrusions in the membrane and this process has surface energy and bending energy costs. Although it is expected that with increasing surface tension and bending rigidity, it should become more difficult to create a protrusion, we find that for a fixed bending rigidity, as the surface tension increases, protrusions are more easily formed. This effect also gives rise to non-trivial dependence of membrane velocity on the surface tension, characterized by a dip and a peak. We explain this unusual phenomenon by studying in detail the interplay of the surface and the bending energy and show that this interplay is responsible for a qualitative shape change of the membrane, which gives rise to the above effect. 
\end{abstract}
\maketitle

\section{Introduction}
\label{sec:intro}
Inside a cell, actin filaments grow and exert polymerization force on the plasma membrane and create membrane protrusions. This process plays an important role in cell motility \cite{review1, review2, review3, review4, review5,farrel2013,banguy2013}. In many experiments, it has been studied how the mechanical properties of the plasma membrane directly affect the formation of membrane protrusion. For example, by artificially increasing the membrane tension in experiments, it has been seen that the rate of protrusion formation goes down, while by decreasing the membrane tension, the rate is found to increase \cite{NilsGauthier2011, DRaucher2000}. Although, these studies indicate that the membrane tension is generally an obstacle to protrusion formation, in \cite{EllenLBatchelder2011}, it was shown that in certain situations, membrane tension can also enhance protrusions by streamlining actin polymerization in one specific direction.

These studies show it is important to understand how the elastic interactions in the membrane influence the mechanism of protrusion formation and in this rapid communication, we address this question in a simple setting. We model the elastic membrane using the Helfrich Hamiltonian, which is the most commonly used model that includes surface and bending energy of the membrane \cite{Helfrich1973, Julian2016, Lawrence2005, Daillant2005, Quemeneur2014, Bossa2018, Fournier2014, Lenz2009, lin2006, peleg2011, mark2010, isaac2013, shlomovitz2011, veksler2007, veksler2009, orly2014}. We describe the membrane using a one dimensional height field, whose time-evolution is governed by the Helfrich Hamiltonian. Any variation in the height costs energy and a flat membrane corresponds to the lowest energy configuration. We consider the membrane being pushed by few growing filaments, which tend to create protrusions in the membrane that cost energy \cite{baumgaertner2010, baumgaertner2012, Sadhu2018, lipowsky94}. As surface tension $\sigma$ or bending rigidity $\kappa$ of the membrane is increased, one would expect that protrusion formation should become more difficult, since the energy cost for creating a protrusion ought to go up monotonically with $\sigma$ and $\kappa$. Surprisingly, we find it is not so. For a fixed value of $\kappa$, there is a significant range of $\sigma$, for which the energy cost actually decreases with $\sigma$, which makes protrusion formation easier. More specifically, the surface energy cost for creating a local protrusion in the membrane increases monotonically with $\sigma$, as expected, but the bending energy cost shows a peak, which in turn gives rise to a peak in the total energy cost. Not only for cell motility, our finding has potential implications for a wide class of systems, where elastic deformation of a membrane is involved. The fact that we have been able to observe this effect in a simple and general model is encouraging and this opens up the exciting possibility of finding it in many different kinds of systems.

\begin{figure}[h!]
\centering
\includegraphics[scale=0.7]{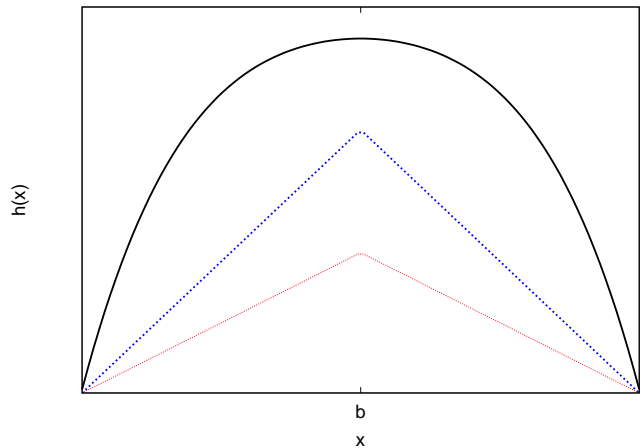}
\caption{Schematic picture of typical membrane shape $h(x)$ around the binding site at $x=b$. Keeping $\kappa$ constant at a moderate or large value, as $\sigma$ is varied, the membrane shape changes qualitatively. The curve at the bottom (top) is for the largest (smallest) $\sigma$ value.}
\label{avg_height_schematic} \end{figure}

To understand the mechanism behind this intriguing effect, we examine the shape of the membrane near the binding site, where the filament is in contact with the membrane, and show that a qualitative change in the shape is responsible for this. In Fig. \ref{avg_height_schematic}, we depict this mechanism. In the limit when both $\sigma$ and $\kappa$ are large, the membrane has very slow spatial variation of its height. In this limit, one can neglect higher order derivatives of height and assume the height profile of a local protrusion is almost linear in space, as shown in Fig. \ref{avg_height_schematic} by the thin red line. For such a local shape of the membrane, the bending energy (which scales as the square of the second derivative of height) has a non-zero value at the binding site. Everywhere else on the membrane near the binding site, the bending energy is negligible. As $\sigma$ decreases slightly but still remains large, the shape of the membrane remains qualitatively the same, but the height gradient magnitude now increases (see the blue dotted line in Fig. \ref{avg_height_schematic}) and the bending energy becomes significantly larger at the binding site. To minimize this energy, as $\sigma$ is lowered further, the membrane shape finally changes qualitatively, and becomes as shown by the black thick line in Fig. \ref{avg_height_schematic}, where the peak gets rounded and the height profile also does not remain linear anymore. While the bending energy now is non-zero even away from the binding site, its variation across the membrane happens more gradually. The bending energy cost to create a local protrusion, which is proportional to the fourth derivative of height, is lower for such a configuration. 

%%%%%%%%%%%%%%%%%%%%%%%%%%%%%%%%%%%%%%%%%%%%%%%%%%%%%%%%%%%
\section{Description of the model}
\label{sec:model}
\begin{figure}[h!]
\centering
\includegraphics[scale=0.8]{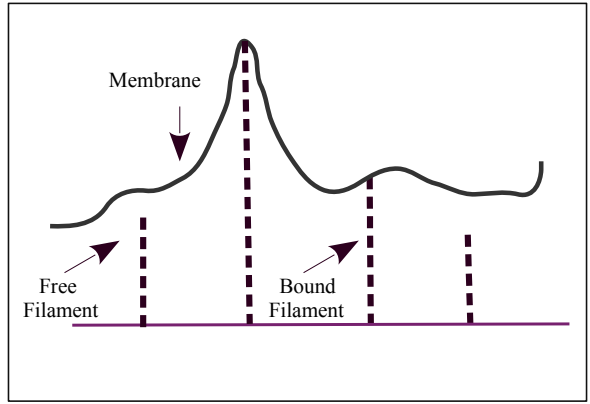}
\caption{Schematic diagram of the model. The thick solid line represents the height profile of the membrane and thick dashed lines represent the bound and free filaments. }
\label{model} \end{figure}

Our system consists of a set of $N$  parallel filaments growing against a membrane as shown in Fig. \ref{model}. The membrane is described by a height profile $\{h_i\}$, defined on a one dimensional lattice of length $L$ with lattice constant unity. We assume periodic boundary condition $h_i = h_{i+L}$ on the lattice. In the absence of any external force, the height profile of the membrane follows an equilibrium dynamics with the Helfrich Hamiltonian, which is standardly used to describe the height fluctuations of a plasma membrane \cite{lin2006,mark2010,isaac2013,shlomovitz2011}. This Hamiltonian consists of surface interaction and bending interaction of the membrane and in our lattice model, it has the form \cite{nelson, lipowsky94, AVolmer1998, baumgaertner2012}
\be
{\cal H}=\sigma \sum_{i=1}^L (h_i')^2 + \kappa \sum_{i=1}^L (h_i'')^2=\sigma \sum_{i=1}^L (h_i-h_{i-1})^2 + \kappa \sum_{i=1}^L (h_{i-1}-2h_i+h_{i+1})^2,
\label{eq:hamiltonian} 
\ee
where $\sigma$ is proportional to the surface tension and $\kappa$ to the bending rigidity. Both of these parameters have dimensions of energy/length$^2$, as follows from the above equation. In Eq. (\ref{eq:hamiltonian}), we have neglected the non-linear terms, which can be justified if the magnitude of height gradient everywhere on the membrane is much less than unity \cite{veksler2007,nirgov2006,Kabaso2011}. The size of a typical eukaryotic cell is about a few tens of micrometers \cite{book2}, which can be compared with the size of our membrane patch. The horizontal distance between two consecutive lattice sites in our model can be taken to be $1/L$ times the size of the membrane. The vertical height difference between two consecutive lattice sites is in the scale of nanometers, set by the size of an actin monomer. Thus, height gradient remains small and the linear approximation for the Hamiltonian in Eq. (\ref{eq:hamiltonian}) remains valid.

It follows from Eq. (\ref{eq:hamiltonian}) that the minimum energy configuration is reached when the membrane is completely flat, {\sl i.e.} $h_i$ the same for all $i$. For a finite temperature, the membrane undergoes thermal fluctuations, as a result of which $h_i$ can increase or decrease by an amount $\delta$. The corresponding energy costs $\mathcal{E}_i ^\pm$ can be obtained from  Eq. (\ref{eq:hamiltonian}) as
$$\mathcal{E}_i^\pm=\sigma \{2 \delta ^2 \mp 2 \delta (h_{i-1}-2h_i+h_{i+1})\} + \kappa \{6 \delta ^2 \pm 2 \delta (h_{i-2}-4h_{i-1} +6h_i-4h_{i+1} + h_{i+2})\}$$
or, in terms of the discrete derivatives, 
\be
\mathcal{E}_i^\pm=\sigma \{2 \delta ^2 \mp 2 \delta h_i''\} + \kappa \{6 \delta^2 \pm 2 \delta h_i''''\}= \Sigma_i^ \pm  + \mathcal{K}_i^\pm.
\label{eq:energy_cost} \ee
Here $\mathcal{E}_i ^ +$ denotes the total energy cost for changing $h_i$ to $h_i + \delta$ and $\Sigma_i^ +$ denotes the surface energy cost and $\mathcal{K}_i^+$ denotes the bending energy cost for the same process. Similarly,  $\mathcal{E}_i ^ -$, $\Sigma_i^ -$, and $\mathcal{K}_i^-$  are the respective energy costs for changing $h_i$ to $h_i -\delta$. We use the Metropolis algorithm to perform the simulations.

One additional constraint that the system must obey is that the height of the membrane at the binding sites should be such that the membrane always stays above the filament tips (as shown in Fig. \ref{model}). Any height fluctuation that brings a binding site at a lower height than the filament tip, is forbidden. As long as this constraint is satisfied, the update rules can be chosen following local detailed balance $R_+ / R_- = e^{-\beta \epsilon}$, where $R_+$ is the rate of a process that has a positive energy cost $\epsilon$ and $R_-$ is the rate of the reverse process.

The filaments are modeled as rigid rod-like polymers, composed of monomers of length $d$ \cite{kolomeisky2015, Sadhu2016, Sadhu2018, Sadhu2019}. In Fig. \ref{model}, we have represented the filaments using dashed lines, each dash denoting a monomer. A (de)polymerization event increases (decreases) the length of a filament by an amount $d$. For the sake of simplicity, we have used $d =\delta$ here [also see Eq. (\ref{eq:energy_cost})]. In Sec. \ref{sup:sec7} of \cite{sup}, we have included our data for the $d\neq\delta$ case. There are two types of filaments in our model: a free filament, which is not in contact with the membrane, and a bound filament, whose tip is in contact with the membrane site \cite{kirone,Sadhu2016,Sadhu2018,Sadhu2019}. The point of contact is referred to as a binding site. For a free filament, polymerization happens with rate $U_0$. For a  bound filament, however, a polymerization process increases the height of the binding site by an amount $d$ and hence there is an energy cost involved in this process. For a positive (negative) energy cost, the bound filament polymerization rate is taken to be $U_0 R_+$  ($U_0 R_-$), while for zero energy cost, the rate is simply $U_0$. The depolymerization rate of the filament is equal to $W_0$ always, as it does not involve any membrane movement. Note that, in the absence of any polymerization force from the filaments, the membrane tends to stay flat and this aspect is somewhat similar to lamellipodial protrusions, rather than filopodial protrusions observed in a cell. In our simulation, each Monte Carlo step consists of $L$ membrane updates and $N$ filament updates. The detailed simulation algorithm has been presented in Sec. \ref{sup:sec1} of \cite{sup}.

%%%%%%%%%%%%%%%%%%%%%%%%%%%%%%%%%%%%%%%%%%%%%%%%%%%%%%%%%%%%%%%
\section{Results} 
\label{sec:results}
We present our results in this section. As mentioned in the Introduction, our main result is the non-monotonic variation of the membrane velocity as a function of $\sigma$. We first present our simulation data showing this effect and then we show how this non-trivial behavior can be explained from detailed measurement of the energy cost and the shape of the membrane. We show our results for a single filament ($N=1$) here. Most of our conclusions remain valid even for the case of multiple filaments, if the filament density is not too high. 

\subsection{Membrane velocity shows a dip and a peak with $\sigma$}
\label{sec:sigma_v}
\begin{figure}[h!]
\centering
\includegraphics[scale=0.7]{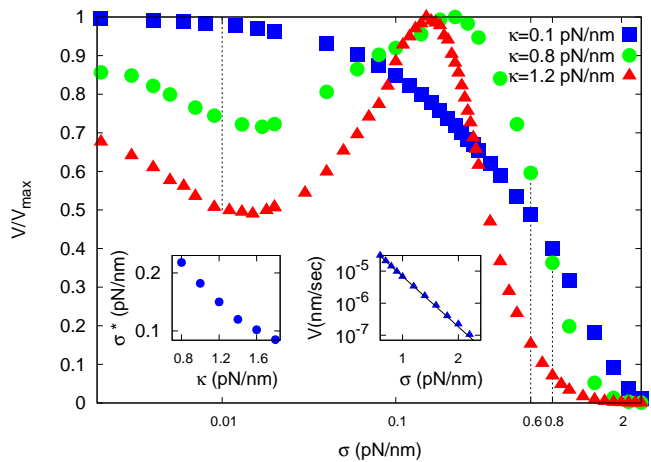}
\caption{$\sigma-V$ curve for different values of $\kappa$.  For small $\kappa$ values, $V$ decreases with $\sigma$, but for moderate or large $\kappa$ values, $V$ shows a minimum and a maximum. We scale each curve by $V_{max}$, which denotes the largest value of $V$ for a particular $V-\sigma$ curve. We have $V_{max} = \num{3.45e-2} nm/sec$, $\num{1.55e-3} nm/sec$ and $\num{2.00e-4} nm/sec$ for $\kappa=0.1 pN/nm$, $0.8 pN/nm$ and $1.2 pN/nm$ respectively. The vertical lines show $\sigma$ values used in Fig. \ref{avg_height}. Left inset: peak position $\sigma^*$ shifts leftward  with $\kappa$. Right inset: Exponential fall of $V$ for large $\sigma$ with $\kappa=1.2 pN/nm$. The decay constant $3.6$ matches closely with the analytical prediction $2\beta \delta ^2$. Here, for all the plots, we use $L=64$. The filament depolymerization rate, $W_0 = 1.4 s^{-1}$ \cite{review1, pollard,kirone}, the free filament polymerization rate, $U_0=2.784 s^{-1}$ \cite{review1,pollard,kirone}, the monomer size is $d=\delta=2.7 nm$ \cite{review1, hansda2014,kirone}, $\beta=1/k_B T$, and we have used $T=300 K$.}
\label{sigma_vs_v} \end{figure}

Pushed by the growing filaments, the membrane develops an average velocity. In Fig. \ref{sigma_vs_v}, we present our data for the variation of membrane velocity $V$  as a function of the surface tension $\sigma$, for a fixed value of bending rigidity $\kappa$. We find that for small $\kappa$, velocity decreases monotonically with $\sigma$, as expected \cite{Sadhu2018}. However, as $\kappa$ is held fixed at a moderate or large value, $V$ shows a rich behavior: starting with a non-zero value at $\sigma=0$, $V$ first decreases with $\sigma$ and reaches a minimum and then it increases to reach a maximum before finally decreasing exponentially for large $\sigma$. While this non-monotonic behavior is in general interesting \cite{Sadhu2018}, the most intriguing observation here is that, there is a range of $\sigma$ for which $V$ grows with $\sigma$. This growth is counter-intuitive because one generally expects that with increasing surface tension, it should become more difficult for the filament to push the membrane. As we show in the following section, this expectation in fact breaks down.

The data in Fig. \ref{sigma_vs_v} is for a fixed $L$ value. We have checked that as $L$ is increased, $V \sim 1/L$, but the peak position does not depend on $L$ for sufficiently large $L$ values. As $L$ becomes small, the peak in the $V-\sigma$ plot shifts to lower $\sigma$ values with decreasing $L$ (Fig. \ref{sigma_vs_v_diff_L} of \cite{sup}).

%%%%%%%%%%%%%%%%%%%%%%%%%%%%%%%%%%%%%%%%%%%%%%%%%%%%%%%%%%%%%%%%%%
\subsection{Bending energy cost shows a peak with $\sigma$ } 
\label{sec:4th_derivative}
\begin{figure}[h!]
\centering
\includegraphics[scale=0.7]{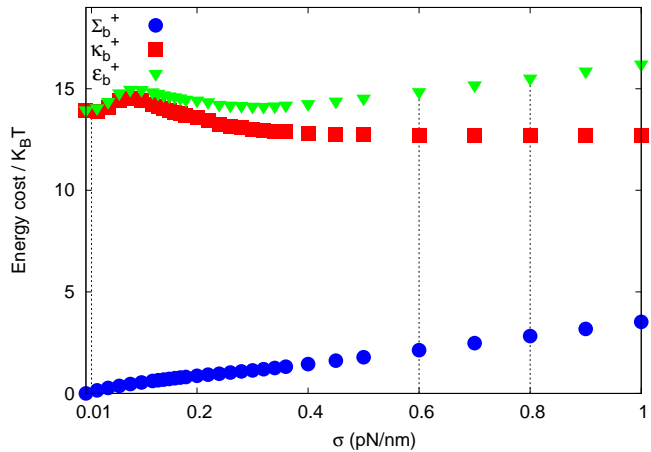}
\caption{Average energy cost for creating a protrusion at the binding site. Surface energy cost $\Sigma_b^+$ increases with $\sigma$ but bending energy cost $\mathcal{K}_b^+$ shows a peak. For moderate or large $\kappa$ values, when $\mathcal{K}_b^+$ is dominant, this gives rise to a peak in total energy cost $\mathcal{E}_b^+$. For very large $\sigma$, we see  $\mathcal{K}_b^+$ saturating and variation in $\mathcal{E}_b^+$ is then controlled by $\Sigma_b^+$ again. These data are for $\kappa=1.2 pN/nm$ and the other simulation parameters are as in Fig. \ref{sigma_vs_v}. The vertical lines correspond to $\sigma$ values used in Fig. \ref{avg_height}.}
\label{sigma_vs_del_e} \end{figure}

From Eq. (\ref{eq:energy_cost}), it follows that the energy cost $\mathcal{E}_b^+$ for creating a protrusion at the binding site $i=b$ can be decomposed into two parts, the surface energy cost $\Sigma_b^+$ and the bending energy cost $\mathcal{K}_b^+$. In Fig. \ref{sigma_vs_del_e}, we plot the steady state average values of $\mathcal{E}_b^+$, $\Sigma_b^+$ and $\mathcal{K}_b^+$ as a function of $\sigma$, for a large value of $\kappa$. While $\Sigma_b^+$ increases with $\sigma$ monotonically, $\mathcal{K}_b^+$ is found to show a peak. Moreover, this peak appears for relatively small $\sigma$ values, when the total energy cost is actually controlled by the bending interaction. As a result, $\mathcal{E}_b^+$ also shows a peak and immediately after the peak, there is a range of $\sigma$ values, for which energy cost decreases with $\sigma$, contrary to the normal intuition. Finally, for very large $\sigma$, when the membrane is almost flat, $\mathcal{K}_b^+$ saturates and $\Sigma_b^+$ increases linearly with $\sigma$ [see Eq. (\ref{eq:energy_cost})] and $\mathcal{E}_b^+$ also shows a linear rise.

\begin{figure}[h!]
\centering
\includegraphics[scale=0.7]{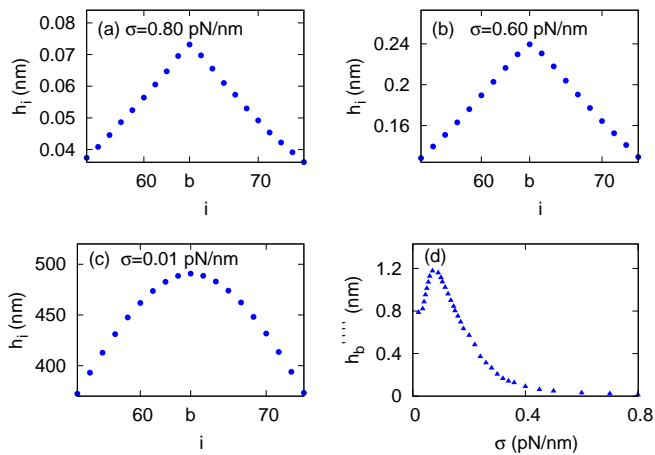}
\caption{(a)-(c): Local shape of the membrane around the binding site for different $\sigma$ with $\kappa=1.2 pN/nm$. (d): Discrete fourth derivative of height at the binding site, $h_b''''=h_{b-2}-4h_{b-1}+h_b-4h_{b+1}+h_{b+2}$, as a function of $\sigma$. Here, $L=64$ and the other simulation parameters are as in Fig. \ref{sigma_vs_v}.} 
\label{avg_height} \end{figure}

Although, it seems quite surprising that energy cost shows non-monotonic variation with $\sigma$, this effect can be very simply explained from the consideration of the membrane shape near the binding site. For very large $\sigma$ and $\kappa$, the membrane remains almost flat and higher order derivatives of height can be ignored and one can assume an almost linear height variation around the binding site. We check this explicitly in our simulations [see Fig. \ref{avg_height}(a)]. Note that, here we only present the height profile near the binding site and far away from this point, height variation does not remain linear anymore, the profile gradually flattens out. However, this part of the membrane does not influence the energy cost near the binding site. In the region around the binding site, where the height profile is linear, the bending energy, which depends on the second derivative of height, has a non-zero value only at the binding site, and zero elsewhere. Keeping $\kappa$ fixed at large value, as we lower $\sigma$,  the qualitative shape of the membrane remains similar, but the magnitude of the height gradient is now larger [Fig. \ref{avg_height}(b)]. This, however, gives rise to a very high bending energy around the binding site, since height gradient changes sharply from a large positive value to a large negative value across the binding site. Therefore, as $\sigma$ is lowered further, such a membrane shape becomes unsustainable and instead a shape shown in Fig. \ref{avg_height}(c) is observed, where the height gradient does not change so sharply but varies more slowly in space. Such a membrane shape stores less bending energy than the one with a linear height variation. In Sec. \ref{sup:sec3} of \cite{sup}, we have also included an approximate analytical explanation of this effect, where we analytically calculated the membrane shape, adapting a simplified description outlined in \cite{nirgov08}. 

With the above picture in mind, it is now easy to understand the variation of the fourth derivative of height (that determines the bending energy required to create a protrusion) at the binding site. In the configuration shown in Fig. \ref{avg_height}(a), the second derivative has a negative value at the binding site and zero elsewhere in its neighborhood. As we lower $\sigma$, the configuration changes to what is shown in Fig. \ref{avg_height}(b) and the magnitude of the second derivative at the binding site  increases, {\sl i.e.} the minimum in $h_i^{\prime \prime}$ is now sharper. This will increase its curvature, {\sl i.e.} the fourth derivative at the binding site. However, as the membrane shape changes to Fig. \ref{avg_height}(c) for even lower $\sigma$, $h_i^{\prime \prime}$ now has a non-zero (negative) value even some distance away from the binding site, since the height profile is not linear anymore. Around the binding site, $h_i^{\prime \prime}$ now varies more gradually and while it still has a minimum at the binding site, the minimum is not as sharp. This corresponds to a lower curvature and the fourth derivative decreases in magnitude. We show this explicitly in Fig. \ref{avg_height}(d). 

However, when $\kappa$ is small, the above effect is absent. Since the bending energy is actually responsible for the change in membrane shape and for small $\kappa$, the bending energy is just not significant enough to bring about this change. In this case, the membrane height profile around the binding site remains linear for all $\sigma$ values and the fourth derivative falls monotonically with $\sigma$ \cite{sup}. As a result, energy cost and velocity varies monotonically with $\sigma$ (see Figs. \ref{shape_small_k} and \ref{kappa_vs_v} in \cite{sup}).  

%---------------------------------------------------------------------------

\subsection{A more quantitative explanation of the full $V-\sigma$ curve} 
\label{sec:exp}

In the previous section, we have offered a simple qualitative explanation of how the energy cost for creating a protrusion shows a peak with $\sigma$, for large values of $\kappa$. This makes it plausible that $V$ can also show a similar peak, since pushing the membrane upward becomes easier as $\sigma$ is increased within a certain range. However, the net membrane velocity results from both upward and downward movements of the membrane and in this section, we present a more quantitative and detailed calculation, that explains the full $V-\sigma$ curve, along with its maximum and minimum.

The membrane being in steady state, its average velocity should be the same at all sites and below, we write down the expression for velocity at the binding site $b$. For large $\kappa$, height gradients at the binding site are small and
hence energy costs $\mathcal{E}_b^ \pm$ are positive [also see Eq. (\ref{eq:energy_cost})], which gives 
\be
V_b=\delta [U_0 p_0 e^{-\beta \mathcal{E}_b^+} + e^{-\beta \mathcal{E}_b^+} - (1-p_0) e^{-\beta \mathcal{E}_b^-}],
\label{eq:vb} \ee
where $p_0$ is the contact probability of the filament tip with the  membrane. The first term in Eq. (\ref{eq:vb}) corresponds to the bound filament polymerization. The second and third terms represent thermal fluctuation of the membrane height at the binding site. Here, we have used the fact that $h_b$ can always increase with rate $e^{-\beta {\mathcal E}_b^+}$, but it can decrease only when the filament is not bound to the membrane.

As explained in Eq. (\ref{eq:energy_cost}), $\mathcal{E}_b^+$ ($\mathcal{E}_b^-$) is the energy cost for increasing (decreasing) $h_b$ by an amount $\delta$, and generally it depends on the local configuration around the binding site. Here, we use the approximation that $\mathcal{E}_b^\pm$ may be replaced by their average values (to keep our notations simpler, we have used the same symbol for the average quantities as well). We have also checked that (data not shown here) as long as $\kappa$ is not too small, the contact probability $p_0$ does not show much variation and remains close to $1/2$, its value for a rigid barrier \cite{Sadhu2018}.

For large $\kappa$ and small $\sigma$, the surface energy cost is much smaller than the bending energy cost and we can replace $\mathcal{E}_b^\pm \approx \mathcal{K}_b^\pm$ and this gives 
\be 
V \approx e^{-6 \beta \delta ^2 \kappa} \delta \{(U_0+1)p_0 e ^{-2 \beta \delta \kappa h_b''''}-2(1-p_0) Sinh{(2\beta \delta \kappa h_b'''')}\}.
\label{eq:vk} \ee

It is easy to see from the above expression that as $h_b^{\prime \prime \prime \prime}$ first increases with $\sigma$, reaches a peak and then decreases, $V$ also initially decreases with $\sigma$, reaches a minimum, and then starts increasing. As $\sigma$ increases further, it is not possible to neglect the elastic energy anymore and Eq. (\ref{eq:vk}) does not remain valid. However, in this case, both $\sigma$ and $\kappa$ are large, and the magnitudes of $h_b^{\prime \prime}$ and $h_b^{\prime \prime \prime \prime}$ are negligible, and we can write
\be 
V \simeq \delta  (U_0+1) p_0 e^{-6\beta \delta ^2 \kappa} e^{-2 \beta \delta ^2  \sigma}, \label{eq:vks}
\ee
where $V$ decreases exponentially with $\sigma$. We verify this from our numerical simulation (Fig. \ref{sigma_vs_v}, right inset). The exponential decay constant numerically observed is $3.6$, which is close to the analytically predicted value $2\beta \delta ^2 \simeq 3.52$. When $\kappa$ is held fixed at a larger value, the exponential decay in Eq. (\ref{eq:vks}) starts at a smaller value of $\sigma$, since the height derivatives become negligible already. This is why the peak position $\sigma^\ast$ shifts towards smaller values as $\kappa$ increases (Fig. \ref{sigma_vs_v}, left inset).

%%%%%%%%%%%%%%%%%%%%%%%%%%%%%%%%%%%%%%%%%%%%%%%%%%%%%%%%%%%%%%%%%%%%%
\section{Discussions}
\label{sec:sum}
Throughout this rapid communication, we have limited our studies to one dimension and it is an important question whether our conclusions remain valid for a two dimensional membrane as well. Generalizing Eqs. (\ref{eq:hamiltonian}) and (\ref{eq:energy_cost}) for the two dimensional case, it can be shown that the bending energy cost is much higher in this case, which slows down the time evolution for moderate or high values of $\kappa$. Although, for small $\kappa$ values we have been able to verify (data not shown here) that $V$ decreases monotonically with $\sigma$, as found in the one dimensional system, higher $\kappa$ values remain numerically inaccessible to us. More research is needed to conclude with certainty if non-monotonic variation of $V$ with $\sigma$ for high $\kappa$ values persists in the two dimensional model as well.

It should be possible to experimentally verify our conclusions. One direct and perhaps the simplest measurement would be to measure the membrane velocity for different values of $\sigma$ and $\kappa$ \cite{marcy2004, baudry2011, theriot2005} and see whether a non-monotonicity as shown in Fig. \ref{sigma_vs_v} can be found. The energy scales involved in our simulations are actually comparable to physical systems. For example, the increase of $V$ with $\sigma$ that we observe for high $\kappa$, corresponds to the surface energy varying in the range of $0.5 $  to $2$ ($ 10^{-19}$ $J$) and the bending energy lies in between $1$ and $3.5$ ($ 10^{-19}$ $J$). These values are within the experimentally observed ranges for flexible membranes \cite{Dimova}. Although, in real systems, many other factors, apart from surface energy-bending energy interplay, are relevant, it would be interesting to see whether this basic signature of a protrusion formation mechanism can still be found.

%%%%%%%%%%%%%%%%%%%%%%%%%%%%%%%%%%%%%%%%%%%%%%%%%%%%%%%%%%%
\section{Acknowledgements} 
We acknowledge useful discussions with P. Pradhan and A. Kundu. SC acknowledges financial support from the Science and Engineering Research Board, India (Grant No. EMR/2016/001663). The computational facility used in this work was provided through the Thematic Unit of Excellence on Computational Materials Science, funded by Nanomission, Department of Science and technology (India).

%%%%%%%%%%%%%%%%%%%%%%%%%%%%%%%%%%%%%%%%%%%%%%%%%%%%%%%%%%%%%% 

\newpage
\section*{\textbf{\large{Supplementary Material: Interplay between surface and bending energy helps membrane protrusion formation}}}
\renewcommand{\theequation}{S-\arabic{equation}}
\setcounter{equation}{0}
\renewcommand{\thefigure}{S-\arabic{figure}}
\setcounter{figure}{0}
\renewcommand{\thesection}{\Roman{section}}
\setcounter{section}{0}

\section{Details of simulations}
\label{sup:sec1}
We perform simulations using kinetic Monte Carlo technique. Any height fluctuation that causes a positive energy cost $\mathcal{E}$, will occur with rate $R_+$ and the reverse movement will occur with rate $R_-$, that satisfies a local detailed balance, i.e., $R_+/R_-=e^{-\beta \mathcal{E}}$. We choose the rates using Metropolis algorithm. For example, if the movement of a membrane site is associated with an energy cost $\mathcal{E}$, then for $ \mathcal{E} > 0$, the rate will be $e^{-\beta  \mathcal{E}}$ and for $ \mathcal{E} \leq 0$, the rate will be unity. We define the relative time scale between the filament dynamics and the membrane dynamics by a parameter ${\cal S}$. In a system that consists of $N$ filaments and $L$ membrane sites, a single Monte Carlo time-step will consists of $N$ filament updates and ${\cal S}$ independent membrane site updates. More specifically, for $N < {\cal S}$, we first choose a filament at random and perform polymerization or depolymerization dynamics. Then we choose ${\cal S}/N$ membrane sites in random sequential order and update them. We repeat this process $N$ times, which completes one Monte Carlo step. Similarly, for $N > {\cal S}$, we first perform $N/{\cal S}$ filament updates and then choose one membrane site at random and update it; repeating this process ${\cal S}$ times completes one Monte Carlo step. Note that a bound filament polymerization also updates the height of the binding site. We start with an initial condition, where the membrane is flat and all filaments are of length $d$ ({\sl i.e.}, each filament consists of only one monomer). The system undergoes time-evolution and after a large number of Monte Carlo steps when the system reaches steady state, we perform our measurements. In the main paper, we have shown data for ${\cal S}/L=1$. In the following section we consider the effect of varying ${\cal S}$. 

\section{Shape of the membrane for small $\kappa$}
\label{sup:sec2}
\begin{figure}[h!]
\centering
\includegraphics[scale=0.7]{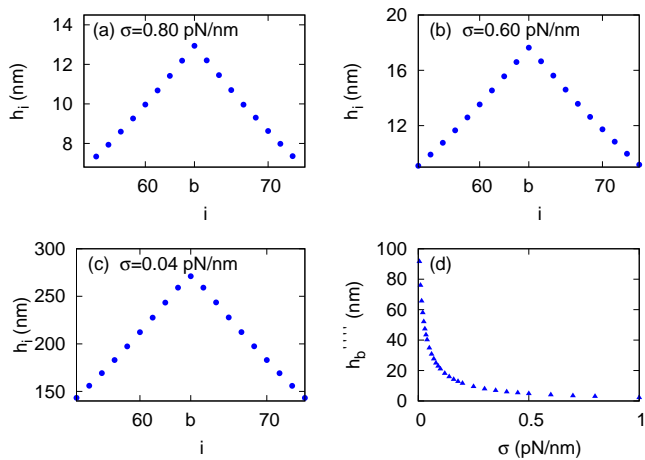}
\caption{(a)-(c): Local shape of the membrane around the binding site for different $\sigma$ with $\kappa=0.01 pN/nm$. (d): Fourth derivative of height at the binding site ($h_b''''$) as a function of $\sigma$. Here we use $L=64$. The other simulation parameters are same as used in the main text.}
\label{shape_small_k} \end{figure}

We show the shape of the membrane for small $\kappa   (=0.01 pN/nm)$ in Fig. \ref{shape_small_k}(a-c). We note that the shape of the membrane remains linear for small as well as large values of $\sigma$. The fourth derivative at the binding site ($h_b''''$) decreases monotonically with $\sigma$ (Fig. \ref{shape_small_k}(d)).

%%%%%%%%%%%%%%%%%%%%%%%%%%%%%%%%%%%%%%%%%%%%%%%%%%%%%%%%%%%%%%%%%%%%%
\section{Analytical calculation of membrane shape}
\label{sup:sec3}
It is possible to give an approximate analytical explanation behind the qualitative change in the membrane protrusion, shown in Figs. 5(a)-(c) of the main paper. The time-evolution equation of a height field of a flexible membrane in presence of an external point force $F$ can be written as \cite{nirgov08}
\be 
\partial_t h(x,t) = \sigma \partial_x^2 h(x,t) - \kappa \partial_x^4 h(x,t) + F \delta (x) + \eta (x,t)
\label{eq:nir}
\ee
where $\eta(x,t)$ is the thermal white noise. Note that in our system the force is being applied by the moving filament tip and we also have the additional constraint that the binding site must stay above the filament tip. A point force with a constant magnitude, as in Eq. \ref{eq:nir}, does not capture these aspects. But this equation is still useful to understand analytically the effect shown in Fig 5 of the main paper.

\begin{figure}[h!]
\centering
\includegraphics[scale=0.7]{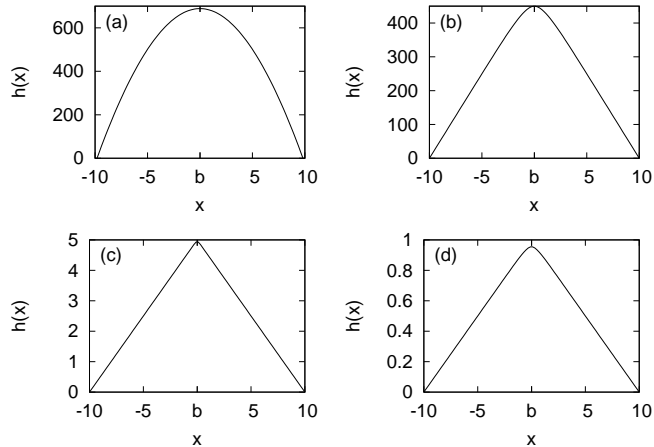}
\caption{Average height profile in Eq. \ref{soln} plotted for different $\sigma$ and $\kappa$. (a): $\sigma =0.001$, $\kappa =1.0$ (b): $\sigma=0.01$, $\kappa= 0.01$ (c): $\sigma=1.0$, $\kappa=0.01$ (d): $\sigma =5.0$, $\kappa=1.0$. We use $F = 1$ here.  }
\label{analytical} \end{figure}

In the long time limit, the left hand side of the Eq. \ref{eq:nir} can be replaced by the velocity $V$ of the membrane. In the co-moving frame, it is possible to analytically solve for the average height at a distance x from the binding site and this solution has the form 
\be 
\langle h(x) \rangle   \sim  -\frac{F}{2 \sigma}\Big( \sqrt{\frac{\kappa}{\sigma}} e^{-\sqrt{\frac{\kappa}{\sigma}} |x|} +|x| \Big).
\label{soln}
\ee
In  Fig. \ref{analytical} we plot the average shape of the membrane for various values of $\sigma$ and $\kappa$ and we see similar change in shape, as
shown in Fig. 5 of the main paper.
%%%%%%%%%%%%%%%%%%%%%%%%%%%%%%%%%%%%%%%%%%%

\section{$\kappa-V$ curves are monotonic for any $\sigma$}
\label{sup:sec4}
\begin{figure}[h!]
\centering
\includegraphics[scale=0.7]{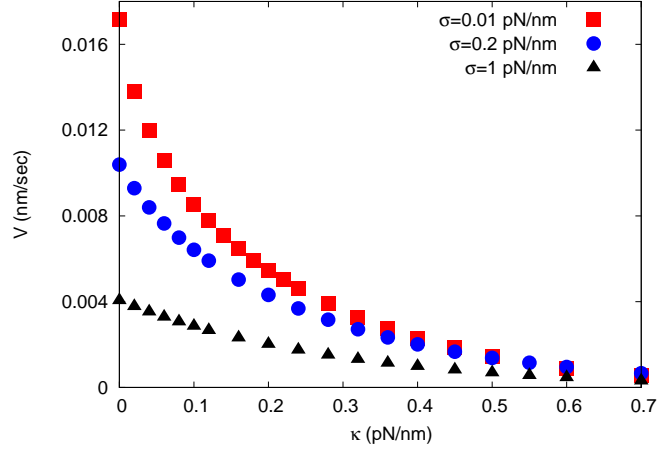}
\caption{$\kappa-V$ curves are monotonic for small and large values of $\sigma$. Here, we use $L=256$ and the other simulation parameters are as in the main text.}
\label{kappa_vs_v} \end{figure}

Although $V$ shows minimum and maximum as a function of $\sigma$, when we plot $V$ against $\kappa$ with $\sigma$ held fixed, we find that $V$ decreases monotonically with $\kappa$. We present our data in Fig. \ref{kappa_vs_v}. As seen in our earlier explanation, the non-monotonic variation in $V$ is associated with non-monotonicity in the energy cost $\mathcal{E}_b^+$ for creating a protrusion at the binding site. Now, as $\kappa$ increases, bending energy cost $\mathcal{K}_b^+$ also goes up. The surface energy cost $\Sigma_b^+$ involves $h_b^{\prime \prime}$ with a negative sign and magnitude of $h_b^{\prime \prime}$ decreases with $\kappa$ (since in the Helfrich Hamiltonian, the bending term scales as square of second derivative which must decrease as $\kappa$ increases). Thus $\Sigma_b^+$ also increases with $\kappa$. As a result, $\mathcal{E}_b^+$ always increases with $\kappa$ and $V$ shows a decrease.  

\section{Faster membrane dynamics lowers $\sigma^*$} 
\label{sup:sec5}
\begin{figure}[h!]
\centering
\includegraphics[scale=0.7]{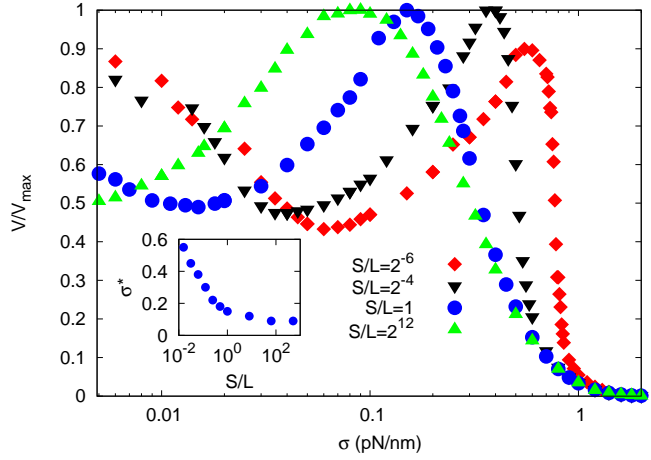}
\caption{$\sigma-V$ curves for different values of ${\cal S}/L$ with large $\kappa$. The peak position as well as the minimum position shifts towards smaller $\sigma$ as ${\cal S}/L$ increases. Inset: The value of $\sigma^*$ decreases with  ${\cal S}/L$. We use $\kappa=1.2 pN/nm$ for the main plot and $\kappa=0.8 pN/nm$ for the inset. For both the plots, we use $L=64$. The other simulation parameters are same as used in the main text.}
\label{sigma_vs_v_diff_s} \end{figure}

The relative time-scale between the thermal fluctuation of the membrane and the filament (de)polymerization affects the qualitative shape of the $V-\sigma$ curve \cite{Sadhu2018,Sadhu2016,Sadhu2019}. So far we considered the case when both these time-scales are same, ${\cal S}/L=1$. In Fig. \ref{sigma_vs_v_diff_s} we show the data for different values of ${\cal S}/L$. It is clear from the data that the curves depend strongly on  ${\cal S}/L$. In particular, the peak position shifts towards smaller values as ${\cal S}/L$ increases (see inset of Fig. \ref{sigma_vs_v_diff_s}). As explained in the previous sub-section, the peak position $\sigma^\ast$ marks the onset of exponential decay of $V$ for large $\sigma$ which corresponds to the situation when the membrane is almost flat and the height derivatives can be neglected. When ${\cal S}/L \ll 1 $, the membrane dynamics is much slower. Thus any local protrusion that is created by the filament at the binding site, persists for a long time before the neighboring sites can adjust their height and flatten the membrane again. Therefore, $\sigma^\ast$ is pushed to larger values. On the other hand, when ${\cal S}/L \gg 1 $, membrane fluctuations are much faster, and in between two filament-moves the membrane gets enough time to relax and become almost flat, which effectively lowers the value of $\sigma^\ast$. 

\section{$\sigma-V$ curve for different values of membrane size (L)}
\label{sup:sec6}
\begin{figure}[h!]
\centering
\includegraphics[scale=0.7]{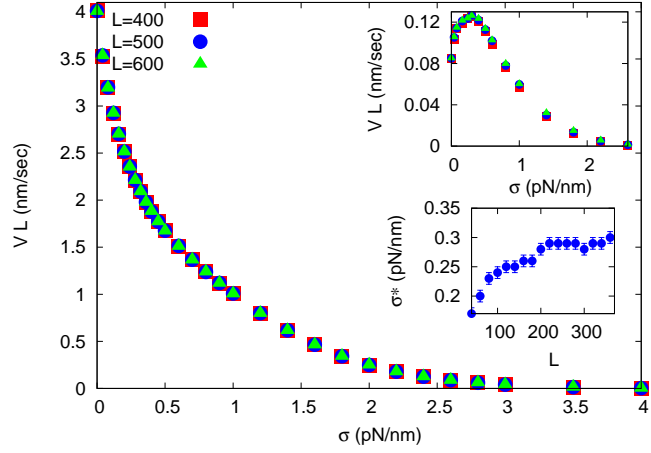}
\caption{$\sigma-V$ curves for different values of $L$. For small value of $\kappa (=0.01pN/nm)$, the curve scales as $1/L$. Top inset: For large value of $\kappa (=0.8 pN/nm)$ also, the $\sigma-V$ curves scales with $1/L$.  Bottom inset: The value of $\sigma^*$ increases with $L$ and saturates for large $L$ values. Here, we use ${\cal S}/L=1$. The other simulation parameters are same as used in the main text.}
\label{sigma_vs_v_diff_L} \end{figure}

We measure $\sigma-V$ curves for different values of $L$. We note that for large $L$, $\sigma-V$ curves collapse for different $L$ if we scale velocity by $1/L$, for both the small and large $\kappa$ (Fig. \ref{sigma_vs_v_diff_L}). Since, the binding site is only the site that is being pushed by the filament and the dynamics of other sites just follow the local detail balance, the overall velocity of the membrane is generated by the drive present at the binding site. This is the reason why the velocity scales as $1/L$. Since, velocity scales with $1/L$ for large $L$, it is expected that the value of $\sigma^*$  will also be independent of $L$. However, for small $L$, membrane velocity does not scale with $1/L$, and $\sigma^*$ is found to shift towards smaller $\sigma$ as $L$ decreases. We plot $\sigma^*$ as a function of $L$ in Fig. \ref{sigma_vs_v_diff_L}, bottom inset and explicitly show this. 

\section{Results for $\delta < d$ case}
\label{sup:sec7}
\begin{figure}[h!]
\centering
\includegraphics[scale=0.7]{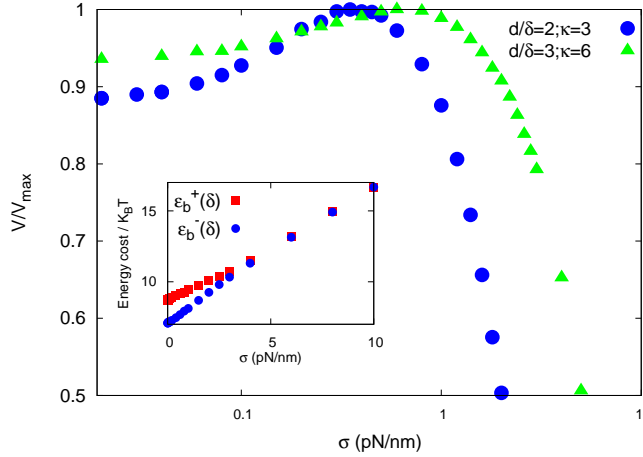}
\caption{Results for $\delta < d$ case. We plot $\sigma-V$ curve for different values of $d/\delta$ with large $\kappa$ (in units of $pN/nm$). We note that similar to the $d=\delta$ case, the membrane velocity shows non-monotonic variation with $\sigma$. Inset: Energy costs $\mathcal{E}_b^\pm (\delta)$ as a function of $\sigma$ for large $\kappa$. We note that the energy cost increases monotonically with $\sigma$. For the inset, we use $d/\delta=2$ and $\kappa=3 pN/nm$ and for all the plots, we use $L=32$ and ${\cal S}/L=1$. The other simulation parameters are same as used in the main text.}
\label{sigma_vs_v_diff_delta} \end{figure}

In this section, we present the data for the case when step length $\delta$ of thermal fluctuation of the membrane is smaller than actin monomer length $d$. In Fig. \ref{sigma_vs_v_diff_delta}, we plot $\sigma-V$ curves for different values of $d/\delta$ when $\kappa$ is held constant at a large value. We find even in this case $V$ shows a peak as a function of $\sigma$, but the minimum of $V$ that precedes the peak for $d=\delta$ case is not observed for $d > \delta$. Comparing Fig. 3 of main paper and Fig. \ref{sigma_vs_v_diff_delta} below at a quantitative level, we also find that the peak of $V$ appears for higher values of $\sigma$ and $\kappa$ as $d/\delta$ is increased.

To explain the origin of the peak in $V$ in this case, let us first consider $d/ \delta=2$ and our argument can be easily generalized to higher values of this ratio. We write down the equation analogous to Eq. 3 of the main paper as
\begin{equation}
V_b=\delta [2 p_0 U_0 e^{-\beta \mathcal{E}_b^+(2\delta)}  + p_1 U_0  e^{-\beta \mathcal{E}_b^+(\delta)} + e^{-\beta \mathcal{E}_b^+(\delta)}  - (1-p_0) e^{-\beta \mathcal{E}_b^-(\delta)}]
\label{supeq:vb}
\end{equation}
where $\mathcal{E}_b^\pm(n\delta)$ is the average energy cost when the binding site height is changed by $\pm n\delta$, with $n=1,2$ and $p_1$ denotes the probability that the filament tip is at a distance $\delta$ from the binding site. Note that in this case the filament polymerization can push the membrane upward by an amount $\delta$ or $2 \delta$. For the parameter range used in our simulations, $\mathcal{E}_b^+(2\delta) >> \mathcal{E}_b^ + (\delta)$ and hence we can neglect the first term in Eq. \ref{supeq:vb}. The inset plot in Fig. \ref{sigma_vs_v_diff_delta} shows the variation of $\mathcal{E}_b^ + (\delta)$ and $\mathcal{E}_b^ - (\delta)$ with $\sigma$ and while these two quantities merge at large $\sigma$, for small $\sigma$ values $\mathcal{E}_b^ - (\delta)$ increases faster. This means the rate of negative displacement of the membrane decreases faster and this explains the rise of $V$ with $\sigma$. For large $\sigma$, membrane is almost flat and $V$ decreases exponentially as in the case of $d=\delta$, discussed in main paper. Our measurement (data not shown here) of average shape of the membrane shows qualitatively similar behavior as shown in Fig. 5 of the main paper. However, unlike $d=\delta$ case, to explain the peak in $V$ here, it is not enough to just consider upward membrane displacement, but the interplay between upward and downward displacements need to be included. 
\end{document}